# Parametric pumping of spin waves by acoustic waves

Pratim Chowdhury, Albrecht Jander and Pallavi Dhagat

School of Electrical Engineering and Computer Science, Oregon State University, Corvallis, USA

**The linear and nonlinear interactions between spin waves (magnons) and acoustic waves (phonons) in magnetostrictive materials provide an exciting opportunity for realizing novel microwave signal processing devices[1–3] and spintronic circuits[4,5]. Here we demonstrate the parametric pumping of spin waves by acoustic waves, the possibility of which has long been theoretically anticipated[6,7] but never experimentally realized. Spin waves propagating in a thin film of yttrium iron garnet (YIG), a magnetostrictive ferrimagnet with low spin and acoustic wave damping, are pumped using an acoustic resonator driven at frequencies near twice the spin wave frequency. The observation of a counter-propagating idler wave and a distinct pump threshold that increases quadratically with frequency non-degeneracy are evidence of a nonlinear parametric pumping process consistent with classical theory. This demonstration of acoustic parametric pumping lays the groundwork for developing new spintronic and microwave signal processing devices based on amplification and manipulation of spin waves by efficient, spatially localized acoustic transducers.**

The interaction between acoustic waves and spin waves includes both linear and nonlinear, parametric effects. The linear coupling between spin waves and acoustic waves, first contemplated theoretically by Kittel[6], has been shown to radiate acoustic waves from resonantly excited ferromagnetic precession[8] and, conversely, excite ferromagnetic resonance[9] and spin waves[4,10,11] in ferromagnetic films upon application of acoustic waves. In the nonlinear coupling regime, parametric excitation of acoustic modes by spin waves, as observed in YIG spheres[12] has been explained by theory developed by Comstock[13,14]. The converse effect, the parametric pumping of spin waves by coherent acoustic waves, however, has not previously been experimentally demonstrated.

The parametric excitation of acoustic waves by spin waves is an important consideration in the design of ferrite-based microwave devices. In most cases, to avoid loss of energy to the acoustic system, the parametric pumping threshold must not be exceeded[15]. In some devices such as frequency selective limiters[1], however, the losses are the basis of device function. The converse pumping of spin waves by acoustic waves could be similarly exploited for technological applications in signal processing, including in spin wave amplifiers, correlators and frequency selective limiters of acoustic signals. In contrast to established methods of parametric pumping of spin waves by electromagnetic waves[16–18], acoustic pumping with piezoelectric transducers promises higher efficiency, localization and ease of integration with micro- and nano-scale circuits. Beyond novel signal processing applications, the recent discovery of spin-caloric effects[19] and acoustically driven spin currents[20,21] provides impetus to the study of magnon-phonon interactions to explain the fundamental processes underlying these phenomena.

Parametric pumping involves the nonlinear interaction between three waves, the signal wave at frequency $f_s$, the pump at frequency $f_p$ and the idler wave at frequency $f_i$. Energy conservation dictates that the three frequencies satisfy the relation

$$f_p = f_s + f_i. \qquad (1)$$

In the present experiments, the pump is a standing acoustic wave that interacts with signal and idler spin waves in a magnetostrictive YIG film.

In the degenerate case where $f_p$ is equal to $2f_s$, the idler frequency is identical to the signal frequency, making it difficult to distinguish the idler from the inevitable electromagnetic feedthrough of the signal wave excitation. As a result, although previous experiments[13] showed modulation of spin wave transmission under the influence of an acoustic pump, they did not convincingly demonstrate the parametric interaction. Here we observe non-degenerate pumping, where the presence of the frequency-shifted, counter-propagating idler as well as a distinct threshold for its appearance provide clear evidence of a nonlinear parametric pumping process.

The device used in our experiments, shown in Figure 1, consists of a thin film piezoelectric transducer fabricated on one side of a 0.5 mm thick gadolinium gallium garnet (GGG) substrate with a ~12 μm thick epitaxial YIG film on the opposite side. The transducer, excites longitudinal acoustic waves, which resonate in the acoustic cavity between the top and bottom free surfaces. The pump frequency is tuned to one of the high-order cavity resonance around 3 GHz to obtain large-amplitude standing acoustic waves in the YIG. The amplitude of the acoustic vibration is controlled by varying the power of the microwave signal applied to the transducer. (See Methods for details on device fabrication and calibration.)

Two microstrip antennas are used for excitation and detection of spin waves in the YIG film, which forms a 1.3 mm wide spin wave waveguide spanning the 8 mm distance between the antennas and passing directly beneath the acoustic transducer. A static magnetic bias field, $H_{BIAS}$, is applied in the film plane, parallel to the waveguide, supporting the propagation of backward volume magnetostatic spin waves between the antennas. Since the pump is orthogonal to the spin waves (see Figure 1(a)), conservation of momentum requires that the idler spin wave propagate counter to the signal wave.

We first examine the propagation of the signal wave through the YIG waveguide, under the influence of the acoustic pump, using a vector network analyzer as illustrated in Figure 2. The pump frequency, $f_p$, is 3022.2 MHz, corresponding to one of the acoustic cavity resonances. The signal spin waves are generated with 1 µW applied to the excitation antenna. The bias field is set to 15.3 mT, the condition at which the transmission of spin waves at $f_s = f_p/2 = 1511.1$ MHz is maximized in the absence of the acoustic pump (see right-most trace in Figure 2).

As the power applied to the acoustic transducer is gradually increased, there is no discernible effect on signal wave transmission until a threshold of about 100 mW is reached. Beyond this threshold, up to 340 mW, the intensity of the transmitted spin wave increases with the acoustic pump power.

We postulate that the accompanying shift in the spin wave spectrum to lower frequencies is associated with a reduction in magnetization due to the pumping of spin waves from the signal wave into modes that do not couple to the receiving antenna. At pump power levels beyond 340 mW, this background of spin waves causes excessive scattering of the signal wave, resulting in the reduction in transmitted power as well.

Next we observe the counter propagating idler spin wave using a circulator at the excitation antenna as illustrated in Figure 3(a). The signal wave is excited using a microwave signal generator swept over a frequency range from $f_s = 1503$ MHz to 1519 MHz. Waves returning to the same antenna are routed to a spectrum analyzer through the circulator. The acoustic pump power is kept at a constant 340 mW.

The detected spectrum is plotted as a function of the signal frequency in Figure 3(b). The main diagonal is the signal frequency, appearing here due to unavoidable reflections from the antenna and electromagnetic feedthrough past the circulator. The parametrically pumped counter-propagating idler wave returning to the transmitting antenna is clearly visible as off-diagonals. The plot is an overlay of the spectra observed for three pump frequencies of 3015.5 MHz, 3022.2 MHz, and 3028.9 MHz (corresponding to adjacent resonant modes of the acoustic cavity). In each case, the frequency relation of equation (1) is maintained. We note that these spectra are not visible when the acoustic cavity is driven off-resonance, eliminating the possibility of electromagnetic interference coupled with a non-linearity in the electronic system being the source of the observed frequencies.

Finally, we examine quantitatively the threshold conditions for parametric pumping for the degenerate as well as non-degenerate cases. For these experiments, the signal frequency was kept constant at $f_s$ =1511.1 MHz while the pump frequency was shifted to different acoustic cavity resonances ($f_p$ =3008.8, 3015.5, 3022.2, 3028.9 and 3035.6 MHz). The observed intensity of the counter-propagating idler wave as a function of the acoustic pump power is plotted in Figure 4. A clear threshold is seen in each case. The threshold increases the further the pump frequency deviates from degeneracy ($\Delta f = f_s - f_p/2 = 50$ kHz is as close as we can get to the degenerate case while still being able to distinguish the idler from the signal frequency). As seen in the inset to Figure 4, the intensity (represented here in amplitude squared, as measured by laser vibrometry[22]) of acoustic waves required to obtain parametric pumping increases quadratically with this frequency offset. We note that the parametric conversion is quite significant, with the intensity of the idler wave reaching nearly 6% of the transmitted wave intensity seen in Figure 2, assuming that the propagation and transducer losses are similar in both cases.

A classical theory for parametric pumping of spin waves was derived by Schlömann, et al.[23] In the most general form, equating the energy pumped into the wave with the damping losses leads to a pumping threshold given by

$$\eta_k^2 = (V_k h_p)^2 - (2\pi \Delta f)^2, \qquad (2)$$

where $h_p$ is the threshold amplitude of a microwave magnetic pumping field, $V_k$ is a coupling factor that depends on the geometry of the device and $\Delta f$ is the offset in pumping frequency from the degenerate case. The spin wave relaxation rate, $\eta_k$, is related to the spin wave linewidth, $\Delta H_k$, by $\eta_k = \gamma \mu_0 \Delta H_k / 2$. Fitting the parabolic dependence on $\Delta f$ to the experimentally determined thresholds (see red trace in inset of Figure 4), we obtain a spin wave linewidth, $\Delta H_k = 85$ A/m (~1 Oe), which is typical of the YIG films used.

In the context of our acoustically pumped device, $h_p$ represents an *effective* magnetic field resulting from the magnetoelastic coupling in the ferromagnetic film. Expanding on the theory of Schlömann[23], Keshtgar et al. recently derived an expression for the coupling of a longitudinal acoustic pump to backward volume magnetostatic waves[24], which, (after accounting for typographical errors) relates the pumping term in equation (2) to the amplitude, $R$, of the acoustic wave as:

$$V_k h_p = \frac{\gamma B_1}{M_s} \frac{2\pi f_p}{c} R. \qquad (3)$$

Here $\gamma$ is the gyromagnetic ratio, $B_1$ the magnetoelastic coefficient, $M_s$ the saturation magnetization and $c$ the longitudinal acoustic wave velocity of the magnetic film. Using equation (3) in equation (2), the threshold

amplitude, $R_c$, for acoustic pumping in the degenerate case ($\Delta f = 0$) is

$$R_c = \frac{\mu_0 \Delta H_k}{2} \frac{M_s}{B_1} \frac{c}{2\pi f_p}. \tag{3}$$

For YIG, we use $\gamma = 2\pi \times 28$ GHz/T, $B_1 = 3.5 \times 10^5$ J/m³, $M_s = 1.4 \times 10^5$ A/m and $c = 7.2$ km/s, giving a theoretical threshold acoustic amplitude of $R_c = 8.1$ pm at $f_p = 3022.2$ MHz. The corresponding experimentally determined threshold of 39 pm is somewhat higher, but on the order of the predicted value. A more comprehensive theoretical model, which takes into account the finite extent of the pump region as well as the non-uniform distribution of acoustic strain through the thickness of the film will be needed to resolve this discrepancy. Nonetheless, these experiments demonstrate that parametric pumping of spin waves by acoustic waves is possible and provide insight into nonlinear phonon-magnon coupling in magnetostrictive materials. Localized and efficient piezoelectric transducers may thus, in the future, be used to generate, modulate and amplify spin wave signals via acoustic pumping in nonlinear microwave signal processing devices and magnonic logic circuits.

**Methods**

*Device fabrication*

The ~12 μm thick epitaxial YIG film was grown on a 0.5 mm thick single-crystal GGG substrate by liquid phase epitaxy. Using a wafer saw, the substrate was subsequently cut into a 1.5 mm wide strip to form the spin wave waveguide. The acoustic transducer was fabricated on the YIG/GGG strip by sputter deposition and shadow masking. The active transducer area of approximately 1.3 mm square is defined by the overlap of 180 nm thick Al electrodes sandwiching the 800 nm thick piezoelectric ZnO. Cu microstrip antennas (both 25 μm wide) were patterned at the ends of two coplanar waveguides on a printed circuit board. The device was taped to this board with the YIG film facing down. The acoustic transducer was connected to a third coplanar waveguide by wire bonding.

*Calibration*

The thickness-mode resonances of the acoustic cavity were determined using a network analyzer to display the absorption spectrum ($S_{11}$) of the acoustic transducer as shown in Figure M1. The acoustic resonances are spaced approximately 6.7 MHz apart, limiting the pump frequency to these discrete values. The amplitude of the acoustic vibration, as controlled by the applied microwave signal power, was calibrated using a heterodyne laser vibrometer[22]. At resonance, the combined effect of the transducer efficiency and the quality factor of the cavity result in standing acoustic waves having an amplitude of 3.3 pm/$\sqrt{mW}$.

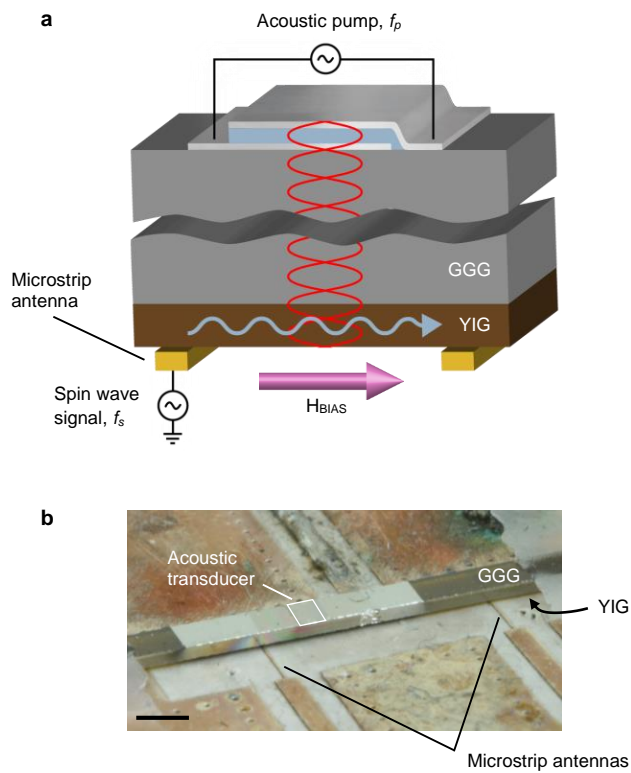

**Figure 1 | Schematic and photograph of the experimental device.** (a) The device is comprised of an acoustic transducer and a YIG spin-wave waveguide on opposite surfaces of a GGG substrate. Microstrip antennas are used to excite and detect spin waves (represented by the wavy blue arrow) in the waveguide. The acoustic transducer consists of a piezoelectric ZnO film sandwiched between Al electrodes. Longitudinal acoustic waves generated by the transducer resonate in the device creating standing waves, as illustrated in red. (b) A photograph of the device. The scale bar is 2 mm.

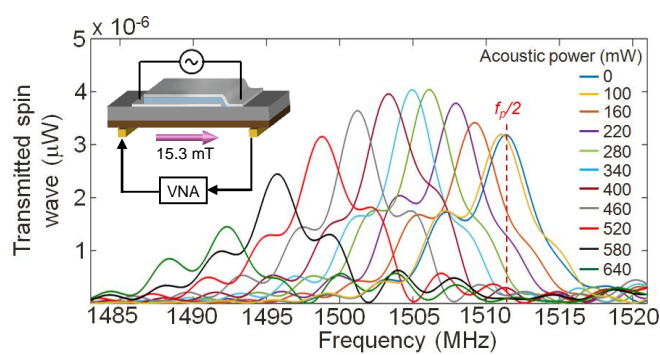

**Figure 2 | Spin wave transmission spectra.** The transmission spectrum of the signal spin waves through the YIG waveguide, as measured by a vector network analyzer (VNA), is shown for various levels of power applied to the acoustic transducer. The schematic, inset in the top left, shows the experimental setup. The signal spin waves are generated with a microwave power of 1 μW applied to the excitation antenna under a bias field of 15.3 mT.

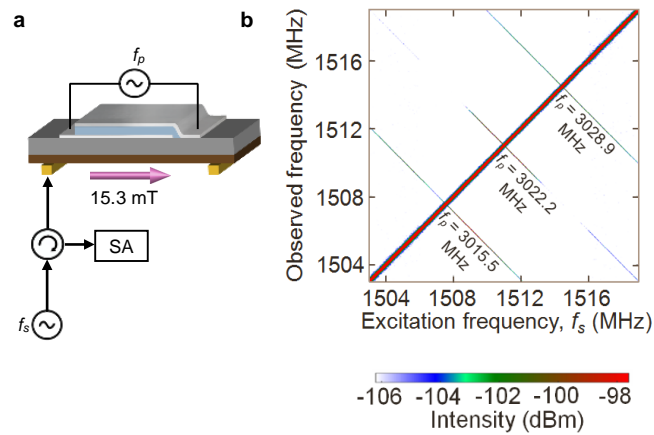

**Figure 3 | Counter-propagating idler spin waves.** (a) Schematic of experimental setup for observing counter-propagating idler spin waves. A spectrum analyzer (SA) connected to a circulator is used to measure the frequency spectrum of waves returning to the excitation antenna. (b) Spectra of waves returning to the excitation antenna versus signal wave frequency. The strong main diagonal is primarily feedthrough of the excitation signal. The off-diagonals show the counter-propagating idler waves that are parametrically excited from the signal wave at different acoustic pump frequencies. The power applied to the acoustic transducer is 340 mW. The signal spin waves are generated under a bias field of 15.3 mT and 1 µW applied to the excitation antenna.

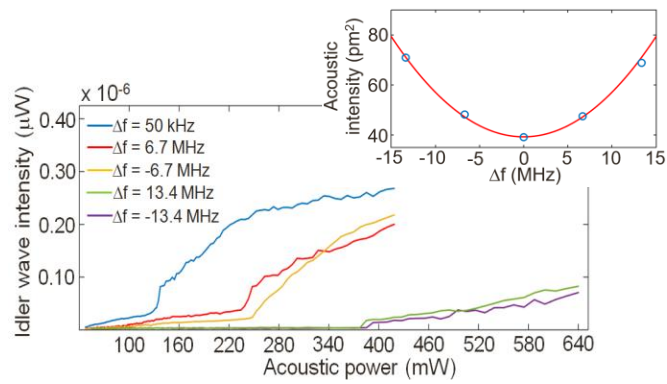

**Figure 4 | Acoustic parametric pumping of spin waves.** Parametrically generated idler wave intensity as a function of acoustic pump power, plotted for different conditions of frequency non-degeneracy. The signal spin waves are excited with 1 µW applied to the excitation antenna. The inset shows the threshold acoustic intensity (in units of amplitude squared) versus frequency offset. The parabolic fit to the data is according to equation (2).

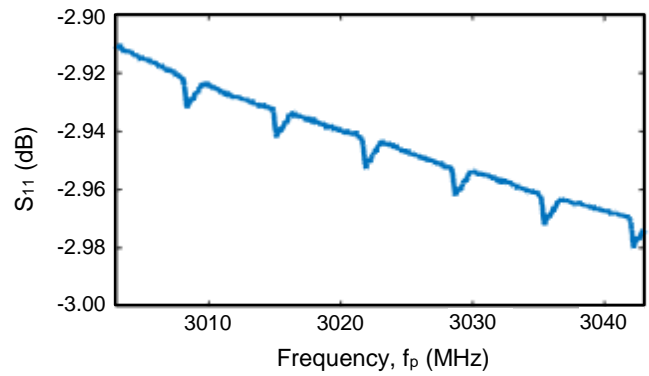

**Figure M1 | Standing wave modes of the acoustic cavity.** The absorption spectrum, $S_{11}(f_p)$ measured at the electrical input to the acoustic transducer, showing the acoustic cavity resonances.

**References**

1. Giarola, A. J., Jackson, D. R., Orth, R. W. & Robbins, W. P. A frequency selective limiter using magnetoelastic instability. *Proc. IEEE* **55,** 593–594 (1967).

2. Robbins, W. P. & Lundstrom, M. S. Magnetoelastic Rayleigh wave convolver. *Appl. Phys. Lett.* **26,** 73–74 (1975).

3. Yao, Z., Wang, Y. E., Keller, S. & Carman, G. P. Bulk acoustic wave-,ediated multiferroic antennas: architecture and performance bound. *IEEE Trans. Antennas Propag.* **63,** 3335–3344 (2015).

4. Cherepov, S. *et al.* Electric-field-induced spin wave generation using multiferroic magnetoelectric cells. *Appl. Phys. Lett.* **104,** 82403 (2014).

5. Dutta, S. *et al.* Non-volatile clocked spin wave interconnect for beyond-CMOS nanomagnet pipelines. *Sci. Rep.* **5,** 9861 (2015).

6. Kittel, C. Interaction of spin waves and ultrasonic waves in ferromagnetic crystals. *Phys. Rev.* **110,** 836–841 (1958).

7. Matthews, H. & Morgenthaler, F. R. Phonon-pumped spin-wave instabilities. *Phys. Rev. Lett.* **13,** 614–616 (1964).

8. Bömmel, H. & Dransfeld, K. Excitation of hypersonic waves by ferromagnetic resonance. *Phys. Rev. Lett.* **3,** 83–84 (1959).

9. Weiler, M. *et al.* Elastically driven ferromagnetic resonance in nickel thin films. *Phys. Rev. Lett.* **106,** 117601 (2011).

10. Pomerantz, M. Excitation of spin-wave resonance by microwave phonos. *Phys. Rev. Lett.* **7,** 312–313 (1961).

11. Gowtham, P. G., Moriyama, T., Ralph, D. C. & Buhrman, R. A. Traveling surface spin-wave resonance spectroscopy using surface acoustic waves. *J. Appl. Phys.* **118,** 233910 (2015).

12. Spencer, E. G. & LeCraw, R. C. Magnetoacoustic resonance in yttrium iron garnet. *Phys. Rev. Lett.* **1,** 241–243 (1958).

13. Comstock, R. L. & Auld, B. A. Parametric coupling of the magnetization and strain in a ferrimagnet. i. parametric excitation of magnetostatic and elastic modes. *J. Appl. Phys.* **34,** 1461–1464 (1963).

14. Comstock, R. L. Parametric coupling of the magnetization and strain in a ferrimagnet. ii. parametric excitation of magnetic and elastic plane waves. *J. Appl. Phys.* **34,** 1465–1468 (1963).

15. Joseph, R. I. & Schlömann, E. Dependence of the phonon-instability threshold for parallel pumping on crystal orientation and magnetic field strength. *J. Appl. Phys.* **41,** 2513–2520 (1970).

16. Kolodin, P. A. *et al.* Amplification of microwave magnetic envelope solitons in thin yttrium iron garnet films by parallel pumping. *Phys. Rev. Lett.* **80,** 1976–1979 (1998).

17. Melkov, G. A. *et al.* Parametric interaction of magnetostatic waves with a nonstationary local pump. *J. Exp. Theor. Phys.* **89,** 1189–1199 (1999).

18. Serga, A. A., Chumak, A. V & Hillebrands, B. YIG magnonics. *J. Phys. D. Appl. Phys.* **43,** 264002 (2010).

19. Bauer, G. E. W., Saitoh, E. & van Wees, B. J. Spin caloritronics. *Nat. Mater.* **11,** 391–399 (2012).

20. Uchida, K. *et al.* Acoustic spin pumping: Direct generation of spin currents from sound waves in Pt/$Y_3Fe_5O_{12}$ hybrid structures. *J. Appl. Phys.* **111,** 53903 (2012).

21. Uchida, K., Qiu, Z., Kikkawa, T. & Saitoh, E. Pure detection of the acoustic spin pumping in Pt/YIG/PZT structures. *Solid State Commun.* **198,** 26–29 (2014).

22. Kokkonen, K., Knuuttila, J. V., Plessky, V. P. & Salomaa, M. M. Phase-sensitive absolute-amplitude measurements of surface waves using heterodyne interferometry. *IEEE Ultrason. Symp. Proc.* **2,** 1145–1148 (2003).

23. Schlömann, E. & Joseph, R. I. Instability of spin waves and magnetostatic modes in a microwave magnetic field applied parallel to the dc field. *J. Appl. Phys.* **32,** 1006–1014 (1961).

24. Keshtgar, H., Zareyan, M. & Bauer, G. E. W. Acoustic parametric pumping of spin waves. *Solid State Commun.* **198,** 30–34 (2014).

**Acknowledgements**

This work was supported in part by the National Science Foundation (Award No. 1414416).

**Author contributions**
P.C. fabricated the devices, performed the measurements and prepared the figures in this manuscript. A.J. and P.D. supervised the work, devised the experiments, interpreted the results and wrote the manuscript.